%% file: main.tex
\documentclass[amsmath,amssymb,aps,prb,reprint,groupedaddress,showpacs]{revtex4-1}

\input{texdef}
\usepackage{amssymb, blindtext, amsmath, lineno, placeins, graphicx, capt-of}
\usepackage{graphicx}
\usepackage{upgreek}
\usepackage{soul}
\usepackage{dcolumn}
\usepackage{bm}

\begin{document}


\title{Analytical and numerical analysis of linear and nonlinear properties of an rf-SQUID based metasurface}


\author{M. M. M\"uller$^1$, B. Maier$^2$, C. Rockstuhl$^{1,3}$, M. Hochbruck$^2$}
\affiliation{$^1$Institute of Theoretical Solid-State Physics, Karlsruhe Institute of Technology (KIT), 76128 Karlsruhe, Germany\\
$^2$Institute for Applied and Numerical Mathematics, Karlsruhe Institute of Technology (KIT), 76128 Karlsruhe, Germany\\
$^3$Institute of Nanotechnology, Karlsruhe Institute of Technology (KIT), 76021 Karlsruhe, Germany}


\date{\today}

\begin{abstract}
We derive a model to describe the interaction of an rf-SQUID
(\textbf{r}adio \textbf{f}requency \textbf{S}uperconducting
\textbf{QU}antum \textbf{I}nterference \textbf{D}evice) based
metasurface with free space electromagnetic waves. The electromagnetic
fields are described on the base of Maxwell's equations. For the
rf-SQUID metasurface we rely on an equivalent circuit model. After a
detailed derivation, we show that the problem that is described by a
system of coupled differential equations is wellposed and, therefore,
has a unique solution. In the small amplitude limit, we provide
analytical expressions for reflection, transmission, and absorption
depending on the frequency. To investigate the nonlinear regime, we
numerically solve the system of coupled differential equations using a finite element
scheme with transparent boundary conditions and the Crank-Nicolson
method. We also provide a rigorous error analysis that shows convergence of the scheme at the expected rates. The simulation results for the adiabatic increase of either the field's amplitude or its frequency show that the metasurface's response in the nonlinear interaction regime exhibits bistable behavior both in transmission and reflection.
\end{abstract}


\maketitle


\section*{Introduction}
\label{sec:Introduction} 
In the last years, researchers spent tremendous efforts in understanding and developing electrodynamic metamaterials that operate at different frequencies from the GHz range up to the visible~\cite{Anlage2011,Jung2014-2,Maas,Cao,Enkrich}.
Metamaterials consist of unit cells that are mostly periodically
arranged in space. These artificially structured materials are
designed to offer control on the propagation of electromagnetic fields
inaccessible with natural materials \cite{Turpin2014}. For that, one
relies frequently on tiny structures inside a host medium to form the
unit cells: the metaatoms. Metaatoms shall assure a strong interaction
of the electromagnetic field with matter. Therefore, resonances are
often exploited. Moreover, controlling the scattering properties of
the individual metaatom is key to tailor the emerging material
properties. For 
a long time, a magnetic response had been looked after but many more properties can be tailored. The metaatoms themselves can be described by purely classical means, e.g., within the context of electrodynamics itself if they are made from ordinary materials such as dielectrics or metals \cite{Kurter2011}, but also by quantum mechanical means if required. That would hold when the metaatom consists of, e.g., a flux qubit as an artifical two-level system \cite{Macha2014}. 

A referential example for a metaatom with a strong magnetic response is the split ring resonator (SRR) \cite{Rock1, Niesler2012, Ricci2006, Ricci2007}. An SRR is a metal ring acting as an inductance with a small gap forming a capacitance, i.e., an LC-circuit. In a natural way, determined by its geometry and material, the SRR has a resonance frequency. However, the downside of using resonant structures made from ordinary metals is (a) a spurious intrinsic absorption that lowers the quality factors and with that the achievable dispersion in the effective properties of the actual metamaterial and (b) their limitation to a fixed resonance frequency upon fabrication \cite{Anlage2011}. 

Both aspects can be mitigated while relying on superconducting materials in the design of metaatoms. First of all, superconductors do not suffer from dissipation \cite{Ricci2005} because they carry current that is not subject to Ohmic resistance due to the bosonic character of their charge carriers \cite{Mangin}. That requires, however, an operational frequency corresponding to an energy that is smaller than the binding energy of the Cooper pairs. This restricts the use of superconductor based metamaterials to the GHz or at most the lower THz frequency range. But superconductors also solve the second aforementioned problem as their properties sensitively depend on the environmental temperature \cite{Chen2010} and magnetic fields they are exposed to \cite{Jung2013,Trepanier2013,Trepanier2017,Ricci2005}. Thus, external parameters have an impact on the intrinsic resonance properties of the metaatoms. 

A further option to tune metamaterials is by exploiting nonlinear effects in the interaction of the electromagnetic wave with the metamaterial. A well understood nonlinear element in the field of superconductivity is the Josephson junction (JJ) \cite{Stewart}. It introduces both nonlinearity into the system and makes use of the low-loss properties of superconducting charge transport. In 2007, it was proposed to put Josephson junctions into the gap of an SRR made of superconducting material and to use these devices as metaatoms \cite{Lazarides2007}. Such structures are called rf-SQUID ring resonators (\textbf{r}adio \textbf{f}requency \textbf{S}uperconducting \textbf{QU}antum \textbf{I}nterference \textbf{D}evice). They are already well investigated in the context of transmission line theory \cite{Butz2013, Jung2014, Jung2013}. Additionally, rf-SQUID rings provide a tunable intrinsic inductivity via an externally applied magnetic field \cite{Jung2013}. Hence, an rf-SQUID is a natural and promising candidate as a building block to create novel, efficient, and tunable metamaterials.

Besides metamaterials as volumetric matter, is has been appreciated that comparable control over electromagnetic fields can be offered by metasurfaces, i.e., thin films made from a monolayer of metaatoms. Then, it is not refraction and diffraction in the bulk media that shall be controlled but rather reflection and transmission from an array \cite{Dolling2006, Rock4, Shalaev2007, Rock3}. In that context, a question of scientific importance concerns the proper description of rf-SQUID based metasurfaces and the exploration of their linear and nonlinear properties. The present contribution develops a comprehensive theoretical framework for that purpose and explores linear and nonlinear properties.

We start by developing an interaction model of free space electromagnetic waves with a thin film loaded with rf-SQUIDs. In the spirit of the metasurface, we assume the spatial extent of the metaatom $L_{\rm MA}$ to be much smaller than the operational wavelength $\lambda$ of the incident wave\cite{OHara2008}, i.e., $L_{\rm MA} \ll \lambda$. Also the thickness $d$ of the metasurface is considered to be much smaller than $\lambda$, i.e., $d \ll \lambda$. Hence, the interaction region can be regarded as infinitesimally thin in the propagation direction of the waves \citep{Caputo2012}. On the one hand, we will describe the dynamics of the system by the continuity of the magnetic field and a jump discontinuity of its first derivative with respect to space, derived from Maxwell's equations \cite{Pfeiffer2013,Selvanayagam2013}. On the other hand, we will use circuit theory and macroscopic quantum effects to describe the inner dynamics of the current and voltage drop inside the rf-SQUID. From these considerations we derive in Section I two coupled differential equations that describe (a) the propagation of the incident field coupled to the rf-SQUIDs and (b) the temporal evolution of the internal dynamics of the rf-SQUID metasurface. 
The wellposedness of our system of equations is proven in Section II. We take this as a justification for the reliability of our approach. The optical response of the metasurface in the linear regime is discussed analytically in Section III. Selected properties of the optical response of the metasurface in the nonlinear regime are discussed numerically in Section IV. For these simulations, we outline an efficient scheme and discuss details of the spatial and temporal discretization of the governing differential equations. This discussion also contains a rigorous error analysis showing error estimates for both discretizations. Finally, we conclude on our work in Section V. 

\section{\label{sec:Derivation} Derivation of the model}
The derivation of a model that describes the interaction of an rf-SQUID ring film with an electromagnetic wave will take into account Maxwell's theory of electrodynamics and circuit theory to express the dynamics in the rings. For the latter we rely on the resistively and capacitively shunted junction model (RCSJ model) of the Josephson junction (JJ) in the rf-SQUID ring \cite{Kleiner, Caputo2012, Caputo2015}. Moreover, we use models of macroscopic quantum effects, such as the Josephson effects and flux quantization. These different aspects are documented in the following subsections. The final purpose of this section is to derive a set of coupled differential equations [cf. \eqref{eq:full_equation}] that describe in a self-consistent manner the evolution of the electromagnetic field and the internal dynamics in the rf-SQUID ring film.  

\subsection{Electrodynamics - Maxwell's equations}\label{chapter: electrodynamics}
\begin{figure}[b]
	\centering
	\includegraphics{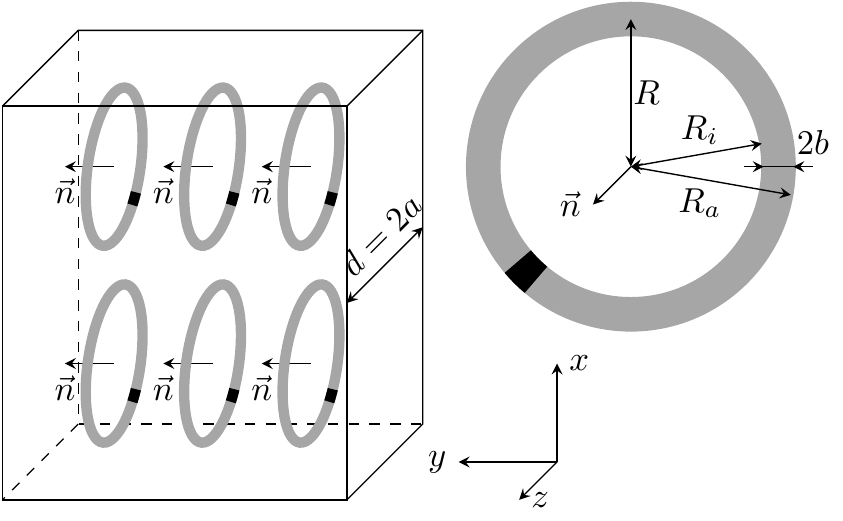}
	\caption{\label{fig:geometry} 
		Schematic view of the geometry and the variables used in the derivation of the model.
	}
\end{figure}
To describe the interaction of an rf-SQUID ring film with electromagnetic fields, we start with Maxwell's equations describing the evolution of electromagnetic fields in time \cite{Jackson},
\begin{subequations}
	\begin{align}
		\vec{\nabla}\times\vec{E}(\vec{r},t) &=-\partial_t\vec{B}(\vec{r},t),\label{eq:maxwell:1}\\
		\vec{\nabla}\times\vec{H}(\vec{r},t) &=\vec{j}(\vec{r},t)+\partial_t\vec{D}(\vec{r},t),\label{eq:maxwell:2}\\
		\vec{\nabla}\cdot\vec{D}(\vec{r},t) &=\rho(\vec{r},t),\label{eq:maxwell:3}\\
		\vec{\nabla}\cdot\vec{B}(\vec{r},t) &=0\label{eq:maxwell:4}.
	\end{align}
\end{subequations}
We set the polarization and the magnetization of the film's host material to zero, since for simplicity, we consider the film to be located in vacuum, such that
\begin{subequations}
	\begin{align}
		\vec{D}(\vec{r},t) &= \varepsilon_0\vec{E}(\vec{r},t),\\
		\vec{B}(\vec{r},t) &= \mu_0\vec{H}(\vec{r},t).
	\end{align}
	\label{eq:constitutiverelations}
\end{subequations}
Differentiating \eqref{eq:maxwell:1} with respect to time and applying the $\operatorname{curl}$ operator to \eqref{eq:maxwell:2} together with \eqref{eq:constitutiverelations} yields
\begin{align}
	\partial_t^2\vec{H}(\vec{r},t)+c^2\vec{\nabla}\times\vec{\nabla}\times\vec{H}(\vec{r},t)=c^2\vec{\nabla}\times\vec{j}(\vec{r},t),\label{wave}
\end{align}
where $c$ is the speed of light in vacuum. This is the governing wave equation that we have to solve to express the dynamics of the electromagnetic field.

As illustrated in Fig. \ref{fig:geometry}, we assume that the film comprising the rf-SQUID rings has a thickness of $d=2a$. Without loss of generality it is located around $z=0$ inside the $x$-$y$-plane, such that $z\in[-a,a]$. This thickness shall be much smaller than the wavelength of the incident light, i.e., $d\ll \lambda$. The orientation of the rings can be arbitrary but we bias our description towards the assumption that the strongest interaction is observed when the rings are upright in the film and the normal vector of the rf-SQUID rings points in $y$-direction. We consider normally incident light which renders our model to be translationally invariant in $x$-$y$-direction, thus $\vec{H}(\vec{r},t)=\vec{H}(z,t)$. Moreover, we assume linear polarization for the magnetic field in $y$-direction. This assures a strong coupling of the magnetic field to the ring at their preferential orientation. 

We start with the evaluation of the left-hand side of \eqref{wave} and have a look at the double $\operatorname{curl}$ of the linearly polarized magnetic field $\vec{H}(\vec{r},t)=H_y(z,t)\hat{e}_y$. It needs a special treatment since the magnetic field $\vec{H}$ is not differentiable twice with respect to space. We make a piecewise ansatz in the three different regions of space (to the left, to the right, and inside the film) and introduce the notation
\begin{align}
\vec{H}^{\Sigma}(z)&:= \vec{H}^-(z)\Theta(-z-a)+\vec{H}^0(z)\Theta(a-z)\Theta(a+z)\nonumber\\
 & \quad+\vec{H}^+(z)\Theta(z-a),
\end{align}
where $\vec{H}$ is forced to be continuous, i.e.,
\begin{align}
\vec{H}^-(-a)=\vec{H}^0(-a),\qquad
\vec{H}^0(a)=\vec{H}^+(a).\label{eq:continuity}
\end{align}
We compute
\begin{align}
		\partial_zH_y^\Sigma &= \left(\partial_zH_y\right)^\Sigma+				\left(H_y^0-H_y^-\right)\delta(z+a)\nonumber\\
 			& \quad +\left(H_y^+-H_y^0\right)\delta(z-a)\nonumber\\
			& =\left(\partial_zH_y\right)^\Sigma,
\end{align}
using the chain rule $\partial_z\Theta(f(z))=\delta(f(z))\partial_zf(z)$ and \eqref{eq:continuity}. Following the same arguments as before, we arrive at
\begin{align}
\partial_z\left(\partial_zH_y\right)^\Sigma &= \left(\partial_z^2H_y\right)^\Sigma+\left(\partial_zH_y^0-\partial_zH_y^-\right)\delta(z+a)\nonumber\\
&\quad +\left(\partial_zH_y^+-\partial_zH_y^0\right)\delta(z-a).
\end{align}
Note that the differences in the brackets do not vanish in general. However, performing the limit $a\rightarrow 0$, we can simplify the expression further and get
\begin{align}
\vec{\nabla}\times\vec{\nabla}\times\vec{H}^\Sigma &= -\partial_z\left(\partial_zH_y\right)^\Sigma\hat{e}_y\label{eq:jump}\\
 &=-\partial_z^2H_y\hat{e}_y+\left(\partial_zH_y^--\partial_zH_y^+\right)\delta(z)\hat{e}_y.\nonumber
\end{align}
For the evaluation of the right-hand side of \eqref{wave}, consider a current density created by a current flowing within a superconducting metal ring. We parametrize the current density in the plane $y=0$ that fully contains the enclosed area of the ring. We call the ring's cross sectional area $A_{\rm c}=\pi(R_{\rm a}-R_{\rm i})^2/4$, where $R_i=R-b$ and $R_a=R+b$ are the inner and outer radius of the ring, respectively. Therefore, one can parametrize the current density's motion as
\begin{equation}
	\vec{j}(\vec{r},t)=\frac{I(t)}{2bA_{\rm c}}\left( \begin{array}{c} z \\ 0 \\-x \end{array}\right)\Theta(R_a-\rho)\Theta(\rho-R_i),\label{current 			density}
\end{equation}
where $\rho=\sqrt{x^2+z^2}$ and $2b$ is the ring's thickness. Due to the cylindrical symmetry, at the position of the origin, where the center of the ring is placed and the interaction with the electromagnetic field takes place, we see that 
\begin{align}
\left(\vec{\nabla}\times\vec{j}(\vec{r},t)\right)_x=\left(\vec{\nabla}\times\vec{j}(\vec{r},t)\right)_z=0.
\end{align} 
The $\operatorname{curl}$ of the current density in \eqref{current density} is therefore given by
\begin{align}
	\vec{\nabla}\times\vec{j}(\vec{r},t) &=\frac{I(t)}{bA_{\rm c}}\Theta(R_a-\rho)\Theta(\rho-R_i)\hat{e}_y\label{current density big}\\
	&\quad + \frac{I(t)}{2bA_{\rm c}}\rho\left[\delta(\rho-R_i)-\delta(R_a-\rho)\right]\hat{e}_y. \nonumber
\end{align}
After expressing $R_a$ and $R_i$ through $R$ and shrinking the ring $(R\rightarrow 0)$, such that it is contained inside the thin film, \eqref{current density big} reads
\begin{align}
	\vec{\nabla}\times\vec{j}(\vec{r},t) &=\frac{I(t)}{bA_{\rm c}}\Theta(b-\rho)\Theta(b+\rho)\hat{e}_y\label{current density big2}\\
	&\quad + \frac{I(t)}{2bA_{\rm c}}\rho\left[\delta(b+\rho)-\delta(b-\rho)\right]\hat{e}_y. \nonumber
\end{align}
In the limit of a vanishing thickness of the ring, such that $b\rightarrow 0$, we notice, that
\begin{subequations}
	\begin{align}
		\lim_{b\to 0}\delta(b+\rho)=\lim_{b\to 0}\delta(b-\rho) &= \delta(\rho),\\
		\lim_{b\to 0}\frac{\Theta(b+\rho)\Theta(b-\rho)}{2b} &= \delta(\rho).
	\end{align}
\end{subequations}
Therefore, only the fist term in \eqref{current density big2} remains and we are left with 
\begin{align}
	\lim_{b\to 0}\vec{\nabla}\times\vec{j}(\vec{r},t) =\frac{2I(t)}{A_{\rm c}}\delta(\rho)\hat{e}_y.\label{eq:cdf}
\end{align}
The Dirac distribution in \eqref{eq:cdf} confines the curl of the current density to $z=0$. Since the model for the entire film is assumed to be translationally invariant in $x$-direction, we omit the confinement to $x=0$ here. This accounts for the existence of other rings along the $x$-direction inside the film. Please note, that the limit in \eqref{eq:cdf} is performed in such a way, that the current density inside the ring remains constant. For an arbitrarily oriented ring's area with unit normal vector $\vec{n}$, from \eqref{wave}, \eqref{eq:jump} and \eqref{eq:cdf}, we can summarize 
\begin{align}
	\partial_t^2\vec{H}-c^2\partial_z^2\vec{H}=\left(\partial_z\vec{H}^+-\partial_z\vec{H}^-+\frac{2I}{A_{\rm c}}\vec{n}\right)c^2\delta(z),\label{eq:totalwave}
\end{align}
linking the current's motion in the ring and the hereby radiated magnetic field $\vec{H}$. Please note, that for $z\neq 0$, we deal with a free wave equation for the magnetic field in the negative and positive half space, respectively,
\begin{equation}
\left(\partial_t^2-c^2\partial_z^2\right)\vec{H}(\vec{r},t)=0.
\end{equation}
Additionally, from \eqref{eq:totalwave}, we obtain a jump condition for the first spatial derivative of the magnetic field at the position of the film $z=0$,
\begin{equation}
\partial_z\vec{H}^+(0,t)-\partial_z\vec{H}^-(0,t)=-\frac{2I(t)}{A_{\rm c}}\vec{n},\label{eq:back-action}
\end{equation}
illustrating that $\vec{H}$ is not differentiable twice at $z=0$. The latter two equations are the most important equations describing the evolution of the field coupled to a film that carries a current. In the next subsection we elaborate on the details of the rf-SQUID ring to express the current in the film that is driven by an external field.

	\subsection{Circuit theory - Kirchhoff's rules}\label{chapter: circuit theory}
We apply the RCSJ model of a Josephson junction to the rf-SQUID ring \cite{Kleiner, Clarke}. It states, that a JJ can be replaced in circuit diagrams by the junction itself (JJ), a shunted capacitor (C), and a shunted resistor (R). Additionally, the ring's loop is taken into account as an inductance connected in series, see Fig. \ref{fig:SQUID}(b). Kirchhoff's nodal rule yields
\begin{align}
	I_{\rm ring}(t) &= I_{\rm C}(t)+I_{\rm R}(t)+I_{\rm JJ}(t) \nonumber\\
		&= C\partial_t U(t)+\frac{U(t)}{R}+I_{\rm JJ}(t).\label{current without josephson}
\end{align}
Kirchhoff's mesh rule yields $\sum_{\rm ring}U(t)=-\partial_t\Phi_{\rm ring}(t)$, where $\Phi_{\rm ring}(t)$ denotes the flux penetrating the ring's enclosed surface acting as the electromotive force. As the voltage drop across all of the three shunted elements is the same, we pick the voltage drop across the JJ for convenience and write 
\begin{align}
	U_{\rm JJ}(t)=-\partial_t\Phi_{\rm ring}(t).\label{voltage without josephson}
\end{align}

	\subsection{Macroscopic quantum effects}\label{chapter: flux quantization}
\begin{figure}[b]
\centering
\includegraphics[scale=0.30]{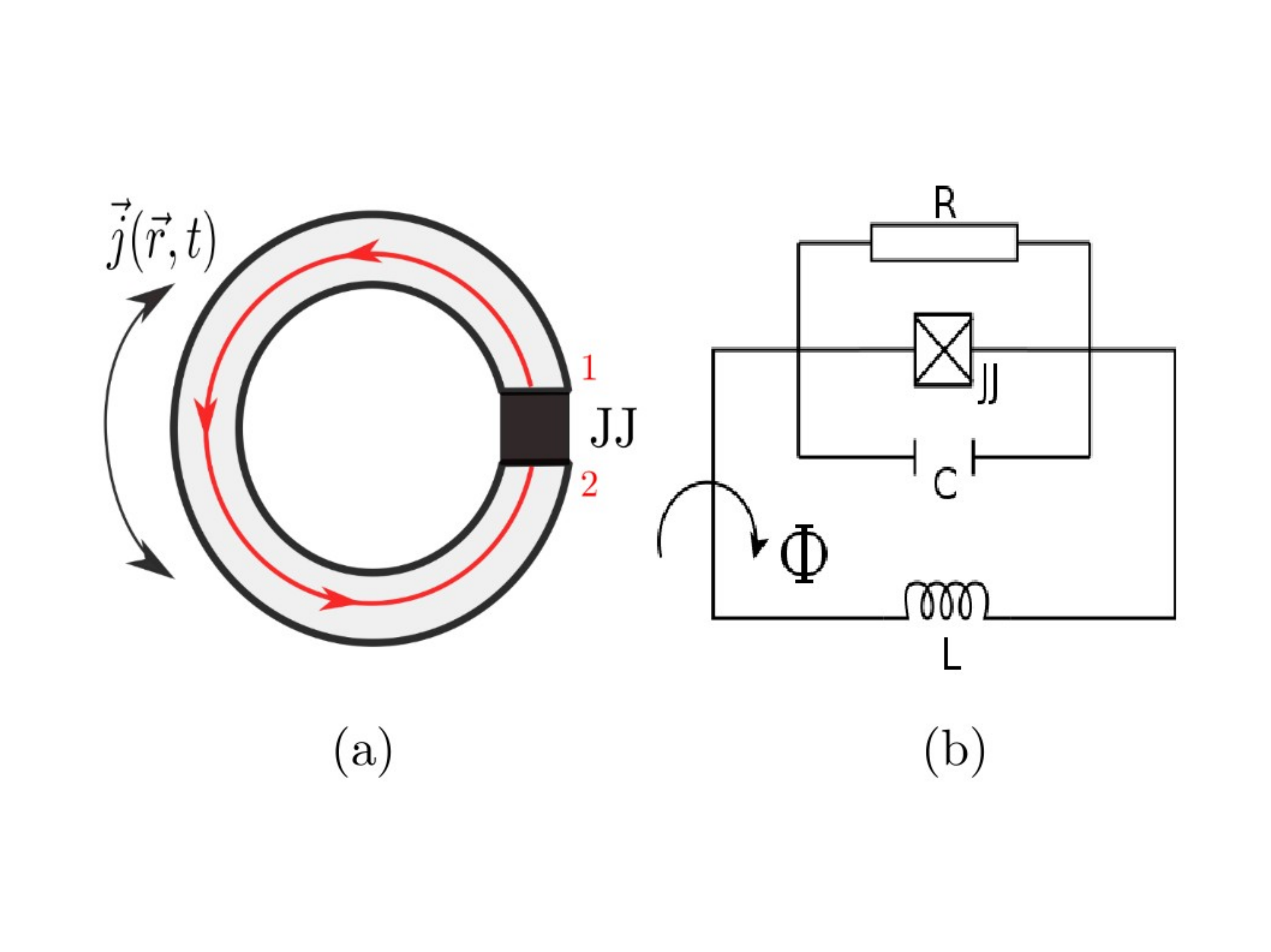}
\caption{(a) Illustration of an rf-SQUID ring with a JJ. (b) shows the equivalent circuit diagram.}
\label{fig:SQUID}
\end{figure}

To link the current through a Josephson junction and the voltage drop across it to the phase difference $\varphi$, we use the gauge invariant definition
\begin{align}
\varphi(t)=\frac{1}{\hbar}\int_{\rm JJ}\vec{p}(t)\cdot d\vec{l}=\frac{1}{\hbar}\int_2^1\vec{p}(t)\cdot d\vec{l}\label{gauge invariant phi}
\end{align}
of the phase difference of the superconducting wave functions on either side of the JJ \cite{Josephson1962}. The integration path $2\rightarrow 1$ in \eqref{gauge invariant phi} refers to Fig. \ref{fig:SQUID}(a). For $\varphi(t)$, Josephson's equations
\begin{subequations}
	\begin{align}
		I_{\rm JJ}(t) &= I_{\rm cr}\sin\varphi(t),\label{josephsoncurrent}\\
		U_{\rm JJ}(t) &=\frac{\hbar}{2e}\partial_t\varphi(t)
	\end{align}
\end{subequations}
hold. Hence, from \eqref{current without josephson} and \eqref{voltage without josephson}, we obtain
\begin{subequations}
	\begin{align}
		I_{\rm ring}(t) &= \frac{C\hbar}{2e}\partial_t^2\varphi(t)+\frac{\hbar}{2eR}\partial_t\varphi(t)+I_{\rm cr}\sin			\varphi(t), \label{current without josephson2} \\
		\frac{\hbar}{2e}\partial_t\varphi(t)&=-\partial_t\Phi_{\rm ring}(t).\label{voltage without josephson2}
	\end{align}
\end{subequations}

Yet another macroscopic quantum effect has to be taken into account, namely the flux quantization in a superconducting loop. This effect occurs when considering a superconducting bulk material device containing a hole \cite{Doll}. We state the general expression of the momentum of a Cooper pair of mass $m=2 m_e$ and charge $q=2 e$ inside a superconductor \cite{Mangin}
\begin{equation}
	\vec{p}(\vec{r},t)=\hbar\vec{\nabla}\phi(\vec{r},t)=2m_e \vec{v}_{\rm q}(\vec{r},t)+2e \vec{A}(\vec{r},t),\label{cooper pair momentum}
\end{equation}
where $\vec{v}_{\rm q}$ denotes the velocity of the Cooper pairs and $\vec{A}$ is the magnetic vector potential, that obeys $\vec{B}=\vec{\nabla}\times\vec{A}$. The term in the middle of \eqref{cooper pair momentum} is generated when applying the momentum operator to the general expression of a condensate's wave function in real space, i.e., $\Psi(\vec{r},t)=\sqrt{n} e^{i\phi(\vec{r},t)}$ with position-independent density distribution $n$. We integrate \eqref{cooper pair momentum} along a closed loop around the superconducting ring 
\begin{align}
	\hbar\oint\vec{\nabla}\phi\cdot d\vec{l} &= \oint\left(2m_e\vec{v}_{\rm q}+2e \vec{A}\right)\cdot d\vec{l}\label{cooper pair momentum int}.
\end{align}
The integration path is chosen such that the distance from the path to the surface of the ring is everywhere larger than the London penetration depth $\lambda_{\rm L}$ of the electromagnetic field into the ring. Then, the integration path only coincides with a current carrying region inside the JJ and $\vec{v}_q(\vec{r},t)\cdot d\vec{l}=0$ holds elsewhere inside the ring. Thus, the current has to be integrated only across the JJ. Please note, that even if the magnetic field penetrates deep into the superconductor and there is no such current-free region, we only have to correct the geometric inductance $L$ in \eqref{screening current} by a kinetic term, i.e., $L\rightarrow L'=L+L_{\rm kinetic}$. By Stokes' theorem, we obtain
\begin{align}
	2\pi m &\approx \frac{1}{\hbar}\int_2^1\left(2m_e\vec{v}_{\rm q}+2e \vec{A}\right)\cdot d\vec{l}+\frac{2e}{\hbar}\int_{A_{\rm r}}\vec{B}\cdot d\vec{F},\nonumber\\
	2\pi m &= \varphi(t)+\frac{2\pi}{\Phi_0}\cdot\Phi(t),\label{fluxquant}
\end{align}
where $\Phi_0=h/2e$ is the flux quantum, $A_{\rm r}=\pi R_{\rm i}^2$ is the enclosed area of the ring, and $m$ an integer. Furthermore, $\Phi(t)$ is the externally applied magnetic flux via a magnetic field and in the first term of the right hand side we used \eqref{gauge invariant phi}. We can see that the total flux $\Phi_{\rm ring}(t)$, which penetrates the ring, consists of the externally applied flux $\Phi(t)=\Phi_{\rm ext}(t)$ and an additional term that describes a screening current $I_{\rm ring}$ in the ring. Its role is to force the enclosed flux onto an integer multiple value of the flux quantum $\Phi_0$, i.e.,
\begin{align}
\Phi_0\cdot m=\frac{\Phi_0}{2\pi}\arcsin\left(\frac{I(t)}{I_{\rm cr}}\right)+\Phi_{\rm ext}(t),\label{phis}
\end{align}
where the current is determined by \eqref{josephsoncurrent}. Equation~\eqref{phis} can be reformulated to
\begin{align}
\Phi_{\rm ring}(t) &=\Phi_{\rm ext}(t)+LI_{\rm ring}(t).\label{screening current}
\end{align}
In the case of the interaction of an electromagnetic wave with the ring, the external flux penetrating the ring's enclosed surface is provided by the magnetic component of the wave, i.e.,
\begin{align}
\Phi_{\rm ext}(t)=\mu_0\int_{A_{\rm r}} (\vec{H}(z=0,t)\cdot\vec{n})dF.\label{fluxmag}
\end{align}
For the sake of brevity and comprehensible readability, we drop the spatial and temporal dependencies from now on, whenever the situation is unambiguous. Hence, from \eqref{current without josephson2}, using \eqref{voltage without josephson2}, \eqref{screening current}, and \eqref{fluxmag}, we arrive at
\begin{align}
\frac{C\hbar}{2e}\partial_t^2\varphi &+\frac{\hbar}{2eR}\partial_t\varphi+I_{\rm cr}\sin\varphi+\frac{\Phi_0}{2\pi L}\varphi=-\frac{\mu_0A_{\rm r}}{L}\vec{H}(0)\cdot\vec{n}.\label{phi unscaled}
\end{align}
Equation \eqref{phi unscaled} is a nonlinear oscillator $\varphi(t)$, that is driven by the magnetic field vector $\vec{H}(z=0,t)$ at the position of the film. Equation \eqref{eq:back-action} describes the back-action of the current in the film on the magnetic field via the jump condition of its first spatial derivative. We know that accelerated charges send out radiation, such that the current can be regarded as the source of the electromagnetic field $\vec{H}(\vec{r},t)$. Equations \eqref{eq:back-action} and \eqref{phi unscaled} constitute the central equations of the interaction model.

	\subsection{Normalization of the model}\label{sec:normalization}	
To investigate their mathematical structure, we boil \eqref{eq:back-action} and \eqref{phi unscaled} down to dimensionless equations by introducing
\begin{align*}
\tilde{\omega} &:=\frac{1}{\sqrt{LC}}, & \tilde{\lambda} &:=\frac{c}{\tilde{\omega}}=c\sqrt{LC},\\
\vec{h}(\vec{r},t) &:=2\pi\frac{\mu_0A_{\rm r}}{\Phi_0}\vec{H}(\vec{r},t), & \alpha &:=\frac{1}{R}\sqrt{\frac{L}{C}},\\
\beta &:=2\pi\frac{LI_{\rm cr}}{\Phi_0}, & \kappa &:=4\pi\frac{A_{\rm r}}{A_{\rm c}}\frac{\mu_0I_{\rm cr}}{\Phi_0}c\sqrt{LC}.
\end{align*}
We use both dimensionless time $\tau:=\tilde{\omega} t$ and space $\xi:=z/\tilde{\lambda}$ variables, where $\tilde{\omega}$ defines a characteristic time scale of the oscillator and $\tilde{\lambda}$ a characteristic length scale of the system. Inserting the above relations, \eqref{eq:back-action} and \eqref{phi unscaled} transform to the dimensionless expressions
\begin{subequations}
	\begin{align}
		\left(\partial_\tau^2-\partial_\xi^2\right)\vec{h} &= 0,\label{wave scaled}\\
		\partial_\xi\vec{h}^+(0)-\partial_\xi\vec{h}^-(0) &= \kappa\vec{n}\left(\vec{h}(0)\cdot\vec{n}+\varphi\right),\label{jump wave scaled}\\
		\partial_\tau^2\varphi+ \alpha\partial_\tau\varphi+\beta\sin\varphi+\varphi &= -\vec{h}(0)\cdot\vec{n}.						\label{phi scaled}
	\end{align}
\end{subequations}
For the sake of simplicity, we will now rename the spatial and temporal coordinates back to the original ones and write $\tau\rightarrow t$ and $\xi\rightarrow z$, both still being dimensionless, i.e.,
\begin{subequations}
\label{eq:full_equation}
\begin{align}
\left(\partial_t^2-\partial_z^2\right)\vec{h} &= 0,\label{wave scaled with t}\\
\partial_z\vec{h}^+(0)-\partial_z\vec{h}^-(0) &= \kappa\vec{n}\left(\vec{h}(0)\cdot\vec{n}+\varphi\right)\label{jump wave scaled with t}\\
\partial_t^2\varphi+ \alpha\partial_t\varphi+\beta\sin\varphi+\varphi &= -\vec{h}(0)\cdot\vec{n}.\label{phi scaled with t}
\end{align}
\end{subequations}
After the physical model of the dynamics in the system has been derived, we now discuss the wellposedness of the problem from a mathematical point of view. This offers a clear indication that the derived system of equations is reasonable.

\section{Wellposedness}\label{sec:WellPosedness}
We now show that \eqref{eq:full_equation} together with initial conditions
\begin{equation}
	\label{eq:initial_cond}
	\begin{aligned}
		\H(0) & = \Ho, & \partial_t \H(0) & = \Hpo, & \text{on } \R,\\
		\varphi(0) & = \varphi_0, & \partial_t \varphi(0) & = \varphi_{\texttt{t},0}, &
	\end{aligned}
\end{equation}
has a unique solution $\H:[0,T] \to H^1(\R)^3 \cap H^2(\R \!\setminus\! \{0\})^3$ and $\varphi:[0,T] \to \R$, where $H^k(\R)$ denotes the Sobolev space of order $k \in \N$.

We prove wellposedness of \eqref{eq:full_equation} with \eqref{eq:initial_cond} using Ref.~\onlinecite{Leib2017}. To keep notation short, we introduce the spaces
\begin{align*}
  \X & = L^2(\R)^3 \times \R, & \V & = H^1(\R)^3 \times \R
\end{align*}
equipped with the respective standard norms. Using the short notation $\u = (\u[1], \u[2]) = (\H, \varphi)$, $\uo = (\Ho, \varphi_0)$ and $\upo = (\Hpo, \varphi_{\texttt{t},0})$, we derive the weak form
\begin{equation} \label{eq:weakForm3D}
	\left\{
	\begin{gathered}
		\text{Find $u :[0,T] \to \V$, such that for all } w \in \V \hfill \\
		\begin{aligned}
			\mt(\partial_{t}^2 \u, w) + \bt(\partial_{t} \u, w) & + \at(\u, w) = \mt(\ft(\u), w), \\
			\u(0) = \uo, \quad & \hphantom{+} \quad \partial_t \u(0) = \upo,
		\end{aligned}
	\end{gathered}
	\right.
\end{equation}
where $\mt : \X \times \X \to \R, \at, \bt : \V \times \V \to \R$ and $\ft : \V \to \X$ are defined by
\begin{align*}
	\mt(\w, \v) & = \int_{\R^3} \w[1] \v[1] \dx + \kappa \w[2] \v[2], \\
  \bt(\w, \v) & = \kappa \alpha \w[2] \v[2], \\
  \at(\w, \v) & = \int_{\R^3} \nabla \w[1] \nabla \v[1] \dx \\
  & \qquad + \kappa ( \w[1](0) \cdot \n + \w[2] ) ( \v[1](0) \cdot \n + \v[2] ), \\
  \ft(\w) & = \begin{pmatrix} 0 \\ - \beta \sin \w[2] \end{pmatrix}.
\end{align*}
Since $\kappa > 0$, $\mt$ is an inner product for $\X$. Moreover $\bt$ is positive semidefinite and continuous with 
\begin{align*}
  \bt(\w, \w) & \geq 0, && \bt(\w, \v) \leq \kappa \alpha \norm{\w}_{\V} \norm{\v}_{\V}, & \w, \v \in \V.
\end{align*}
Furthermore, by Gauss's theorem $\at$ is continuous. It is also symmetric and satisfies a Garding inequality, i.e.,
\begin{align*}
  \at(\w, \v) & \leq C \norm{\w}_{\V} \norm{\v}_{\V}, & \w, \v \in \V,\\
  \at(\w, \w) + c_G \mt(\w, \w) & \geq c_G \norm{\w}_{\V}^{2}, & \w \in \V\hphantom{,}
\end{align*}
for all $c_G > \max\{1,\kappa^{-1}\}$. Finally the right-hand side $\ft$ is Lipschitz continuous with constant $\kappa \beta$, i.e.,
\begin{align*}
  \norm{\ft(\w) - \ft(\v)}_{\X} & \leq \kappa \beta \norm{\w - \v}_{\V}, & \w, \v \in \V.
\end{align*}
Therefore, Theorem~3.3 in Ref.~\onlinecite{Leib2017} yields the existence of a unique solution $\u \in C^2(0,T;\X) \cap C^1(0,T;\V)$ of \eqref{eq:full_equation} with \eqref{eq:initial_cond} for initial values $\uo \in \bigl(H^2(\R^3\!\setminus\!S)  \times \R\bigr) \cap \V$ satisfying the jump condition \eqref{jump wave scaled with t} and $\upo \in \V$ (even if $\uo$ is a bistable point of the potential).

We want to emphasize that an analogous proof holds true without the assumption of the film being translationally invariant with respect to the $x$-$y$-plane, i.e., $\H$ and $\varphi$ being also functions of $x$ and $y$. This situation is relevant if the metasurface shall show a position dependent response to encode further functionalities. 

\section{Analytical treatment in linear approximation}
First, we consider a special case of \eqref{eq:full_equation} and calculate reflection and transmission from the film in the linear regime, where we assume that the effective interaction potential can be accurately described by a parabola. As the one-dimensional case suffices to describe the reflection and transmission coefficients of the film, we continue to assume that the film is placed in the $x$-$y$-plane. We further assume the electromagnetic field and the film to be translationally invariant with respect to the $x$-$y$-plane and calculate the reflection and transmission coefficient of an incident plane wave that is $y$-polarized with its magnetic field. 

We make a small-amplitude ansatz to linearize the differential equations
\begin{subequations}
	\begin{align}
		\vec{h}(z,t) & =\vec{h}_s+\delta\vec{h}(z,t),\label{h ansatz}\\
		\varphi(t) & = \varphi_s+\delta\varphi(t),\label{phi ansatz}
	\end{align}
	\label{eq:linear}%
\end{subequations}
where $\vec{h}_s$ and $\varphi_s$ are the static components of the magnetic field and the phase difference. For $\varphi_s=0$, we obtain
\begin{align}
	\left(\partial_t^2+\alpha\partial_t +\omega_0^2\right)\delta\varphi = -\delta\vec{h}(0)\cdot\vec{n},\label{driven HO}
\end{align}
where we used $\omega_0=\sqrt{1+\beta}$. Analogously to the calculations in Ref.~\onlinecite{Caputo2015}, we proceed with a time-harmonic ansatz to investigate the model in frequency space
\begin{subequations}
	\begin{align}
		\delta\vec{h}(z,t) & =\int d\omega\vec{f}(z,\omega)\exp\left(-i\omega t\right),\label{eq:fourierh}\\
		\delta\varphi(t) & =\int d\omega\phi(\omega)\exp\left(-i\omega t\right).\label{eq:fourierphi}
	\end{align}
\end{subequations}
Equation \eqref{driven HO} yields
\begin{align}
	\left(\omega^2+i\alpha\omega-\omega_0^2\right)\phi(\omega) = \vec{f}(0,\omega)\cdot\vec{n}.\label{eq:HOfourier}
\end{align}
We can plug $\phi(\omega)$ from \eqref{eq:HOfourier} into the jump condition \eqref{jump wave scaled with t} and as a result we eliminate one equation, i.e.,
\begin{align}
\partial_z\vec{f}^+(0,\omega)-\partial_z\vec{f}^-(0,\omega) &=-\kappa\vec{n}\left(\vec{f}(0,\omega)\cdot\vec{n}\right)M(\omega),\label{eq:wavewithM}
\end{align}
where we defined 
\begin{align}
M(\omega)=\frac{\omega^2+i\alpha\omega-\beta}{\omega_0^2-\omega^2-i\alpha\omega}=\frac{1}{\omega_0^2-\omega^2-i\alpha\omega}-1.
\end{align}

	\subsection{Reflection and transmission coefficients}
\begin{figure}[t]
\centering
	\includegraphics[scale=0.32]{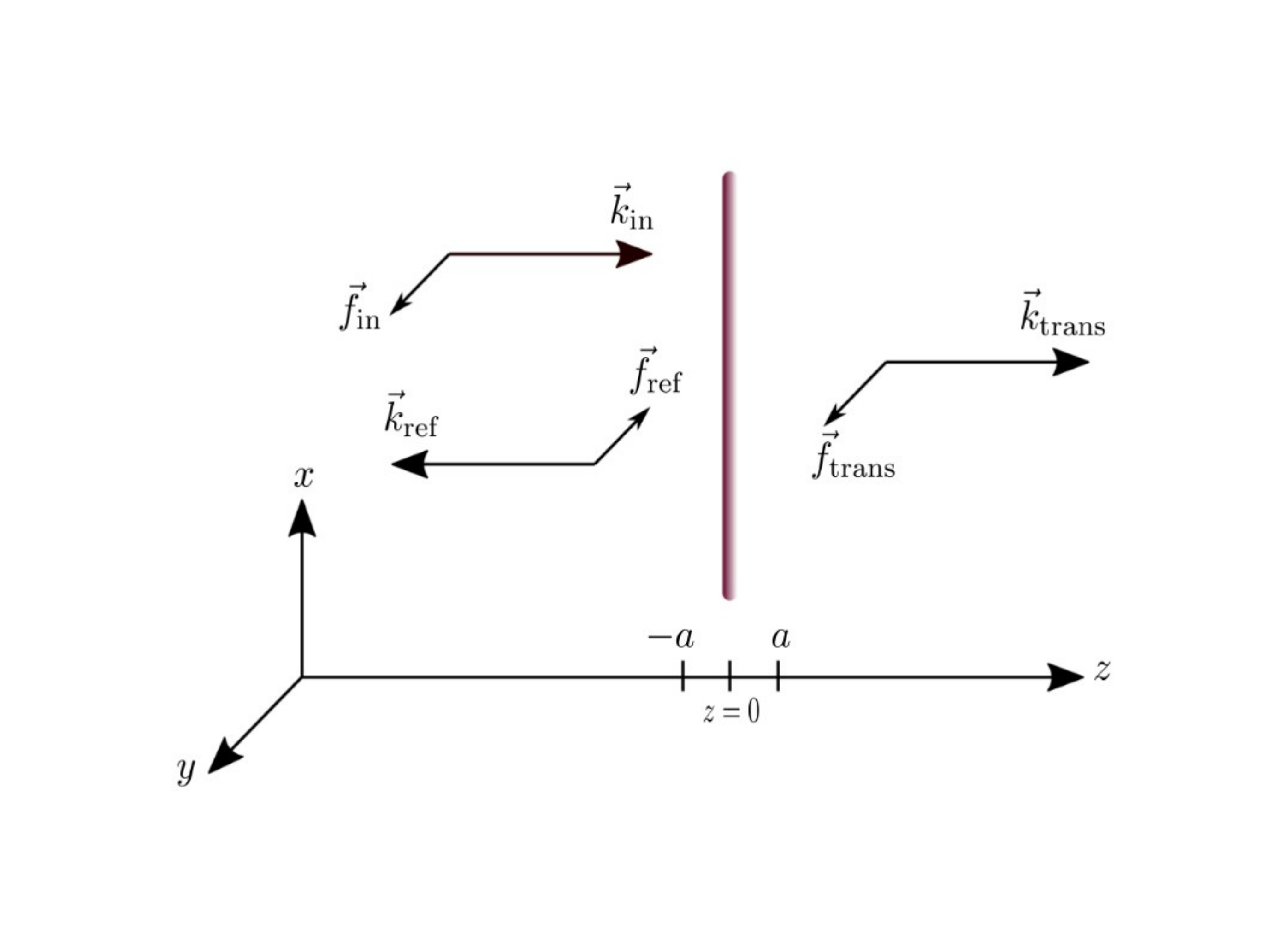}
\caption{\label{fig:zansatz} Graphical illustration of the ansatz for the magnetic field amplitudes in \eqref{eq:ansatzf}.}
\end{figure}
\begin{figure}[b]
	\centering
	\begin{subfigure}[t]{\columnwidth}
		\centering
		\includegraphics{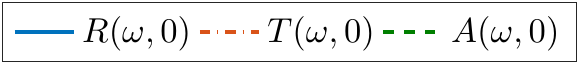}
	\end{subfigure}
	\begin{subfigure}[t]{0.5\columnwidth}
		\centering
		\includegraphics{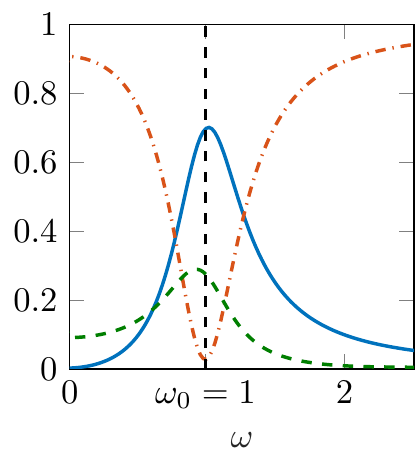}
		\caption{}
	\end{subfigure}%
	\begin{subfigure}[t]{0.5\columnwidth}
		\centering
		\includegraphics{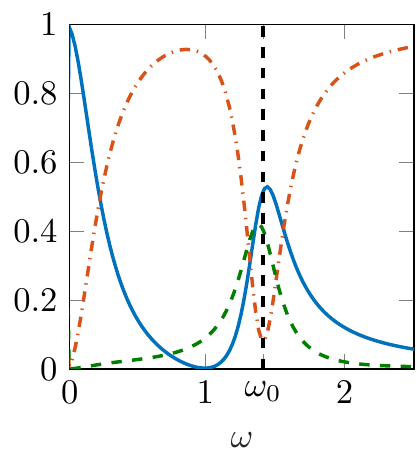}
		\caption{}
	\end{subfigure}
	\caption{\label{fig:RandT} Reflection and transmission coefficients, and absorption functions for $\beta=0$ (a) and $\beta=1$ (b). We chose $\alpha=0.1$, $\kappa=1$.}
\end{figure}

For the spatial dependence of the magnetic field, we also make a harmonic ansatz in the two half-spaces $z<0$ and $z>0$ and impose continuity of the magnetic field at the position of the film $z=0$, see Fig. \ref{fig:zansatz}, i.e.,
\begin{align}
		\vec{f}(z,\omega) &= \begin{cases} \vec{f}_{\rm in}(\omega) e^{i\omega z}+\vec{f}_{\rm ref}(\omega) e^{-i\omega z} & \mbox{if  } z<0,\\
\vec{f}_{\rm trans}(\omega) e^{i\omega z} & \mbox{if  } z>0,\end{cases}\label{eq:ansatzf}\\
\vec{f}_{\rm in}(\omega) &+ \vec{f}_{\rm ref}(\omega) = \vec{f}_{\rm trans}(\omega) \hspace{46pt}\text{at } z=0.\label{eq:singlevalue}
\end{align}
Note that due to the normalization of the model and the propagation direction of the electromagnetic wave along the $z$-axis, it holds $\omega(\vec{k})=k_{\rm z}$. From \eqref{eq:wavewithM} and using the ansatz in \eqref{eq:ansatzf}, we find
\begin{align}
\vec{f}_{\rm trans} &(\omega)-\vec{f}_{\rm in}(\omega)+\vec{f}_{\rm ref}(\omega) = \frac{i\kappa}{\omega}\vec{n}(\vec{f}(0,\omega)\cdot\vec{n})M(\omega).\label{jump condition}
\end{align}
Using \eqref{eq:singlevalue} as well, we obtain
\begin{subequations}
\begin{align}
\vec{f}_{\rm ref}(\omega) &=\frac{i\kappa}{2\omega}M(\omega)\vec{n}(\vec{f}_{\rm trans}(\omega)\cdot\vec{n})\label{eq:refinproject},\\
\vec{f}_{\rm trans}(\omega) &=\vec{f}_{\rm in}(\omega)+\frac{i\kappa}{2\omega}M(\omega)\vec{n}(\vec{f}_{\rm trans}(\omega)\cdot\vec{n}).\label{eq:transinproject}
\end{align}
\end{subequations}
Our goal is to express both the reflected wave $\vec{f}_{\rm ref}$ and the transmitted wave $\vec{f}_{\rm trans}$ through the incoming wave $\vec{f}_{\rm in}$ only. On that account, we project \eqref{eq:transinproject} onto $\vec{n}$ and obtain
\begin{align}
\vec{f}_{\rm trans}(\omega)\cdot\vec{n} &=\frac{\vec{f}_{\rm in}(\omega)\cdot\vec{n}}{1-\frac{i\kappa}{2\omega} M(\omega)}.\label{projecttransform}
\end{align}
As desired, using \eqref{projecttransform}, we can write \eqref{eq:refinproject} and \eqref{eq:transinproject} as functions of the incoming field amplitude $\vec{f}_{\rm in}$ only. We obtain
\begin{align}
\vec{f}_{\rm ref}(\omega) &=i\frac{\kappa M(\omega)}{2\omega-i\kappa M(\omega)}(\vec{f}_{\rm in}(\omega)\cdot\vec{n})\vec{n},\label{fref}\\
\vec{f}_{\rm trans}(\omega) &=\vec{f}_{\rm in}(\omega)+i\frac{\kappa M(\omega)}{2\omega-i\kappa M(\omega)}(\vec{f}_{\rm in}(\omega)\cdot\vec{n})\vec{n}.\label{ftrans}
\end{align}	
For the following, we make assumptions concerning the geometry of the problem. Assuming, that $\theta$ is the inclination angle of the ring's normal vector $\vec{n}$ with respect to the incoming field amplitude $\vec{f}_{\rm in}$, i.e., $\vec{f}_{\rm in}\cdot\vec{n}=|\vec{f}_{\rm in}|\cos\theta$, we find the reflection coefficient $R(\omega,\theta)$ and the transmission coefficient $T(\omega,\theta)$ according to
\begin{subequations}
\begin{align}
R(\omega,\theta):&=\frac{|\vec{f}_{\rm ref}(\omega)|^2}{|\vec{f}_{\rm in}(\omega)|^2}=\frac{\kappa^2}{4\omega^2}\cdot\frac{|M(\omega)|^2}{\text{CD}(\omega)}\cos^2\theta,\label{eq:Rcoeff}\\
T(\omega,\theta):&=\frac{|\vec{f}_{\rm trans}(\omega)|^2}{|\vec{f}_{\rm in}(\omega)|^2}\nonumber\\
 &= 1-R(\omega,\theta)-\frac{\kappa}{\omega}\cdot\frac{\mathfrak{Im}(M(\omega))}{\text{CD}(\omega)}\cos^2\theta,\label{eq:Tcoeff}
\end{align}
\end{subequations}
where we defined the "common denominator" as
\begin{align}
\text{CD}(\omega)=1+\frac{\kappa^2}{4\omega^2}|M(\omega)|^2+\frac{\kappa}{\omega}\mathfrak{Im}(M(\omega)).
\end{align}
Since the absorption function $A(\omega,\theta)$ has to fulfill the energy conservation relation $A+R+T=1$, we find by comparison
\begin{align}
A(\omega,\theta)=\frac{\kappa}{\omega}\cdot\frac{\mathfrak{Im}(M(\omega))}{\text{CD}(\omega)}\cos^2\theta.\label{eq:absorption}
\end{align}
For $\theta=0$, Fig. \ref{fig:RandT} shows the reflection and transmission coefficients, as well as the absorption function for the two cases $\beta=0$ (a) and $\beta=1$ (b). Both cases show the resonance at the resonance frequency $\omega_0=\sqrt{1+\beta}$ of the rings. However, for a notably large current through the Josephson junction $(\beta > 0)$, in the DC limit the film acts as a reflector. This can be explained taking into account self-induction of the rf-SQUID ring loop for small frequencies.

\newcommand{\nthpoint}{1}

\section{\label{sec:Nonlinear} Simulations in the Nonlinear Interaction Regime}
Up to this point, the effects have been computed analytically in the linear interaction regime after having performed linear approximations in \eqref{eq:linear}. We next consider nonlinear effects by numerical simulations. As soon as the amplitude of the incoming magnetic field $\vec{h}_{\rm inc}$ exceeds a critical value, the trigonometric expressions in the equations of motion of our model can no longer be replaced by the linear term of their Taylor expansion.

\subsection{Wellposedness on a bounded domain}
In order to introduce a spatial discretization, we restrict the computational domain to a bounded subdomain $\Ol \coloneqq [\ell_1, \ell_2] \subset \R$ for some $\ell_1 < 0 < \ell_2$. This leads to the following simplified model problem, where the magnetic field $\H$ and the phase $\varphi$ satisfy
\begin{subequations}
	\label{eq:system1D}
	\begin{align}
		(\partial_{t}^2 - \partial_{z}^2) \vec{h} = 0, && \hspace{-8.5ex} [0,T] \times \Ol \setminus \{0\}, & \\
		\partial_{t}^2 \varphi + \alpha \partial_{t} \varphi + \varphi + \beta \sin \varphi & = - \vec{h}(0) \cdot \vec{n}, & \hspace{-10ex} [0,T], &
	\end{align}
	together with the jump condition at the interface $z = 0$
	\begin{align}
		\label{eq:system1D:jump} 
		\kappa \vec{n} \left( h(0) \cdot \vec{n} + \varphi \right) = \partial_z \vec{h}^+(0) - & \partial_z \vec{h}^-(0), & [0,T]. &
	\end{align}
	Following the approach of Ref.~\onlinecite{Grot2009}, we introduce exact transparent boundary conditions
	\begin{equation}
		\label{eq:transpBC}
		\partial_{z} \vec{h}(\ell_1) = \partial_{t} \vec{h}(\ell_1), ~ \partial_{z} \vec{h}(\ell_2) = -\partial_{t} \vec{h}(\ell_2), ~~[0,T].
	\end{equation}
	With these boundary conditions, the solution of the reduced system coincides with the restriction to the domain $\Ol$ of the solution of the original system \eqref{eq:full_equation} with \eqref{eq:initial_cond} if the support of both $\Ho$ and $\Hpo$ is contained in $\Ol$. Therefore, the reduced system yields the same reflection and transmission coefficients as the problem considered on the full space. Finally, we impose again the initial conditions \eqref{eq:initial_cond}, but with $\Ol$ instead of $\R$.

\end{subequations}
The wellposedness of the reduced system is shown analogously to the approach for the general setting. First we introduce the spaces
\begin{align}
	\label{eq:spaces1D}
  \Xo & = L^2(\Ol)^3 \times \R, & \Vo & = H^1(\Ol)^3 \times \R
\end{align}
equipped with the standard norms. The bilinear forms $\mo : \Xo \times \Xo \to \R, \ao : \Vo \times \Vo \to \R$ and the right-hand side $\fo : \Vo \to \Xo$ are defined as before, but with $\Ol$ instead of $\R$. Only the bilinear form $\bo : \Vo \times \Vo \to \R$ changes significantly:
\begin{align*}
	\bo(\w, \v) = & \kappa \alpha \w[2] \v[2] \!+ \w[1](\ell_1) \v[1](\ell_1) + \w[1](\ell_2) \v[1](\ell_2),
\end{align*}
but as all bilinear forms and the right-hand side have the same properties as before, Theorem~3.3 in Ref.~\onlinecite{Leib2017} again yields the existence of a unique solution $u \in C^2(0,T;\Xo) \cap C^1(0,T;\Vo)$ of \eqref{eq:system1D}.

\subsection{Space discretization}
We discretize in space using finite elements on a grid $\Olh$ of $\Ol$. In order to resolve the jump condition \eqref{eq:system1D:jump} correctly, we require $0$ to be a grid point. We denote the maximal length of the intervals in $\Olh$ by $h$. We further introduce the space $\Pk(\Olh)$, consisting of piecewise polynomials of degree at most $k \in \N$ in every interval in $\Olh$, and the space $\Voh \coloneqq \Pk(\Olh)^3 \times \R$.
\begin{equation} \label{eq:weakForm1Dh}
	\left\{
	\begin{gathered}
		\text{Find $\uh \in C^1(0,T;\Voh)$, such that for all } \vh \in \Voh \hfill \\
		\begin{aligned}
			\moh(\partial_{t}^2 \uh, \vh) + & \bo(\partial_{t} \uh,\vh) \\ & + \aoh(\uh,\vh) = \moh(\fo(\uh),\vh), \\
			\uh(0) = \uho, & \hphantom{+} \quad \quad \partial_t \uh(0) = \uhpo,
		\end{aligned}
	\end{gathered}
	\right.
\end{equation}
where the initial values $\uho$ and $\uhpo$ discrete versions of their continuous counterparts. The discrete bilinear forms $\moh, \aoh: \Voh \times \Voh \to \R$ are approximations of $\mo$ and $\ao$, where the integrals are replaced by a quadrature rule of order at least $k^2$. Therefore, the discrete bilinear forms coincide with their continuous counterparts on $\Voh \times \Voh$. Hence, they satisfy the same assumptions and we get from Theorem~3.6 in Ref.~\onlinecite{Leib2017} the following semi-discrete error estimate.

\textit{Theorem (semi-discrete error estimate):}
For the exact solution $\u = (\u[1],\u[2])^T \in C^2(0,T;W^{k,2}(\Ol) \times \R) \cap C^1(0,T;W^{k+1,2}(\Ol) \times \R)$ of the continuous problem and the discrete solution $\uh \in C^2(0,T;\Voh)$ of \eqref{eq:weakForm1Dh}, the following estimate holds for all $t \in [0,T]$
\begin{align}
	& \norm{\uh(t) - u(t)}_{\Vo} + \norm{\partial_t \uh(t) - \partial_t u(t)}_{\Xo} \nonumber\\
	& \quad\! \leq C e^{(\frac{1}{2} + \kappa\beta)t} (1+t) \Bigl( \norm{\uho - \uo}_{\Vo} + \norm{\uhpo - \upo}_{\Xo} \nonumber\\
	& \qquad\! h^k \bigl( \norm{\u[1]}_{\infty, k+1} + \norm{\partial_{t} \u[1]}_{\infty, k+1} + \norm{\partial_{t}^2 \u[1]}_{\infty, k} \bigr) \! \Bigr). \label{eq:errorEstimateSemi}
\end{align}

\subsection{Full discretization}
We use the Crank-Nicolson scheme for the time discretization of \eqref{eq:weakForm1Dh}. First, we define $\Ah, \Bh:\Voh \to \Voh$ via
\begin{equation*}
	\moh(\Ah \uh, \vh) = \aoh(\uh, v), \quad \moh(\Bh \uh, \vh) = \bo(\uh, \vh)
\end{equation*}
for all $\uh, \vh \in \Voh$. The Crank-Nicolson scheme with time step $\tau$ is then given by
\begin{equation}
	\label{eq:CrankNicolson}
	\begin{aligned}
		\begin{pmatrix} \uh^{n+1} \\ \uhp^{n+1} \end{pmatrix} \!=\! \begin{pmatrix} \uhn \\ \uhpn \end{pmatrix} & - \frac{\tau}{2} \begin{pmatrix} 0 & -\Id \\ \Ah & \Bh \end{pmatrix} \!\! \biggl( \!\! \begin{pmatrix} \uh^{n+1} \\ \uhp^{n+1} \end{pmatrix} + \begin{pmatrix} \uhn \\ \uhpn \end{pmatrix} \!\! \biggr) \\
		& + \frac{\tau}{2} \begin{pmatrix} 0 \\ \fo(\uh^{n+1}) + \fo(\uhn) \end{pmatrix}.
	\end{aligned}
\end{equation}
From Corollary~3.7 in Ref.~\onlinecite{Leib2017} we get the following result.

\textit{Theorem (fully discrete error estimate):}
For the exact solution $\u = (\u[1],\u[2])^T \in C^4(0,T;\Xo) \cap C^3(0,T;\Vo)$ and the numerical approximation $\uhn \in \Voh$ obtained by the Crank-Nicolson scheme \eqref{eq:CrankNicolson}, the following error estimate holds for all $\tn = n \tau \leq T$
\begin{equation*}
  \begin{aligned}
		e^n & = \norm{\uhn - u(\tn)}_{\Vo} + \norm{\uhpn - \partial_t u(\tn)}_{\Xo} \\
		& \leq C e^{(\frac{1}{2} + \frac{\kappa\beta}{1-\kappa\beta\tau})\tn} (1+\tn) \Bigl( \norm{\uho - \uo}_{\Vo} \\
		& \quad\! + \norm{\uhpo - \upo}_{\Xo} + \tau^2 \bigl( \norm{\partial_{t}^3 \u}_{\infty, \Vo} + \norm{\partial_{t}^4 \u}_{\infty, \Xo} \bigr) \\
		& \quad\! + h^k \bigl( \norm{\u[1]}_{\infty, k+1} + \norm{\partial_{t} \u[1]}_{\infty, k+1} + \norm{\partial_{t}^2 \u[1]}_{\infty, k} \bigr) \Bigr).
  \end{aligned}
\end{equation*}

\subsection{Validation}
\begin{figure}[b]
	\centering
	\begin{subfigure}[t]{0.51\columnwidth}
		\centering
		\includegraphics{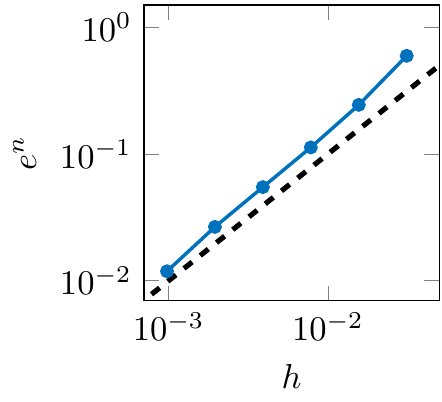}
		\caption{}
	\end{subfigure}%
	\begin{subfigure}[t]{0.51\columnwidth}
		\centering
		\includegraphics{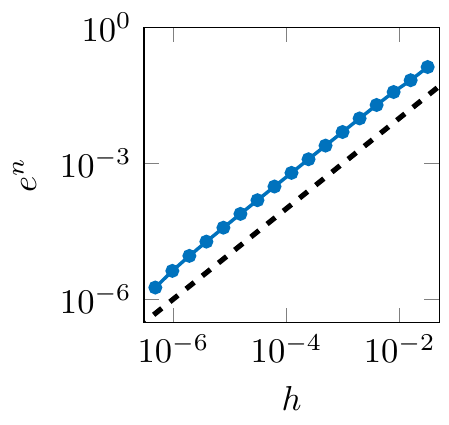}
		\caption{}
	\end{subfigure}
	\caption{\label{fig:error} Error against spatial resolution $h$. \newline
		(a): with b.c. \eqref{eq:transpBC} (no inflow, fixed domain) \newline
		(b): with b.c. \eqref{eq:transpBCinflow} (with inflow, domain $(-h,h)$) \newline
		dashed line indicates first order in $h$
	}
\end{figure}

In Fig.~\ref{fig:error}(a), the error estimate \eqref{eq:errorEstimateSemi} is numerically confirmed for $\alpha = 0, \beta = 3, \kappa = 1$ and $k = 1$. For the initial values, we chose a Gaussian-modulated sinusoidal pulse of the form
\begin{align*}
	\uho(z) & = \begin{pmatrix} -\exp\bigl(-400(z+\frac{1}{2})^2\bigr) \sin\bigl(5 (z+\frac{1}{2})\bigr) \hat{e}_y \\ 0 \end{pmatrix}, \\
	\uhpo(z) & = \frac{d}{dz} \uho(z).
\end{align*}
for $z \in [-\frac{1}{2},\frac{1}{32}] = \Olh$. Since the exact solution is unknown, we computed a reference solution on a finer grid. As predicted, we see linear convergence in the spatial resolution for the error measured in the energy norm.

Since it is not possible to have any inflow with the boundary conditions \eqref{eq:transpBC}, the support of the initial values yields a lower bound for the size of the spatial domain. However, for the reflection and transmission coefficients it is sufficient to know the magnetic field at any pair of points $\varepsilon_1 < 0 < \varepsilon_2$ arbitrarily close to the film, i.e., $-\varepsilon_1, \varepsilon_2 \ll 1$. Therefore, we adapt the boundary condition at $\ell_1$ in order to allow an incident wave $\Hin:[0,T] \to \R$ entering the computational domain $\Ol$ from the left side. Using d'Alembert's formula, $\Hin$ is uniquely defined by the initial values $\Ho$ and $\Hpo$. So in the following, we replace \eqref{eq:transpBC} by the new boundary conditions
\begin{equation}
  \begin{aligned}
		\label{eq:transpBCinflow}
 		\partial_{z} \vec{h}(\ell_1) & = \partial_{t} \vec{h}(\ell_1) - 2 \partial_{t} \Hin, & [0,T], & \\
		\partial_{z} \vec{h}(\ell_2) & = -\partial_{t} \vec{h}(\ell_2), & [0,T]. &
	\end{aligned}
\end{equation}
Although not covered by our analysis, numerical experiments also show first order convergence, as can be seen in Fig.~\ref{fig:error}(b).

The benefit of these boundary conditions is the possibility to drastically reduce the computational domain. In fact, the choice $-\ell_1 = \ell_2 = h$ means that the grid $\Olh$ contains only the two intervals $(-h,0)$ and $(0,h)$. Therefore, the numerical effort for the spatial discretization is completely independent of the spatial resolution.

As the Crank-Nicolson scheme is unconditionally stable, one can even keep the number of time steps constant. Therefore, there is no dependency between spatial resolution and the computational effort. So the maximal computable spatial resolution is solely restricted by the machine epsilon, as the condition number of the resolvent $\begin{pmatrix} \Id & -\frac{\tau}{2}\Id \\ \frac{\tau}{2}\Ah & \Id + \frac{\tau}{2}\Bh \end{pmatrix}$ is growing proportionally to $h^{-2}$.

\subsection{Simulation results}
To investigate the nonlinear effects, we increase the amplitude of the incoming light's magnetic field component and observe the excited phase difference $\varphi$ as well as the reflected and transmitted field amplitude. We apply two qualitatively different types of sources in the simulation setup.
\begin{enumerate}
\item One option is to sweep the amplitude $h_{\rm inc}$ of the incoming magnetic field at a fixed driving frequency $\omega(t)=\omega_{\rm D}$. The corresponding magnetic field for $t\in[0,T]$ has the form
\begin{align}
	\qquad \vec{h}_{\rm inc}(z,t) &= h_{\rm inc}(t)e^{-i\omega_{\rm D}(t-z)}\hat{e}_{\rm y},\\ 
	h_{\rm inc}(t) &= \frac{h_{\rm inc, max}-h_{\rm inc, min}}{T}\cdot t+h_{\rm inc, min},\label{hsweep}
\end{align}
where $h_{\rm inc,min}$ is still part of the linear interaction regime, but $h_{\rm inc,max}$ is not.
\item Another way to observe nonlinear effects is to sweep the frequency of the incoming light at fixed amplitude. In the linear interaction regime, the system provides its maximum amplitude response of $\varphi$ at resonance frequency $\omega_{\rm D}=\omega_0$. This is not necessarily the case, when we go to the nonlinear interaction regime. When we perform a frequency sweep of the incoming light at fixed amplitude, the incoming magnetic field will be of the form
\begin{align}
	\vec{h}_{\rm inc}(z,t) &= h_{\rm inc}e^{-i\omega(t)(t-z)}\hat{e}_{\rm y},\\
	\omega(t) &= \frac{\omega_{\rm max}-\omega_{\rm min}}{T}\cdot t+\omega_{\rm min}.\label{omegasweep}
\end{align}
We will choose the setting in such a way, that at some time, the incoming field is in resonance with the structure, i.e., $\omega_{\rm min}<\omega_0<\omega_{\rm max}$ holds.
\end{enumerate} 

\newcommand{\markrepeat}{6}

\begin{figure}[b]
	\centering
	\begin{subfigure}[t]{\columnwidth}
	  \centering
		\includegraphics{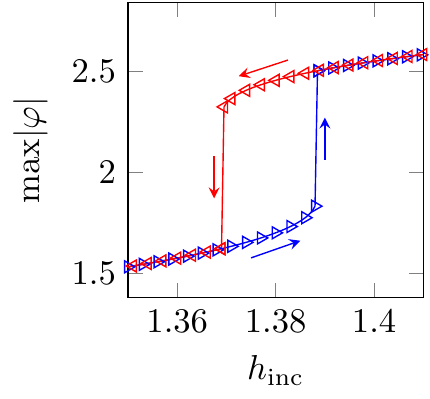}
		\caption{stationary state amplitude}
	\end{subfigure}
	\begin{subfigure}[t]{0.5\columnwidth}
	  \centering
		\includegraphics{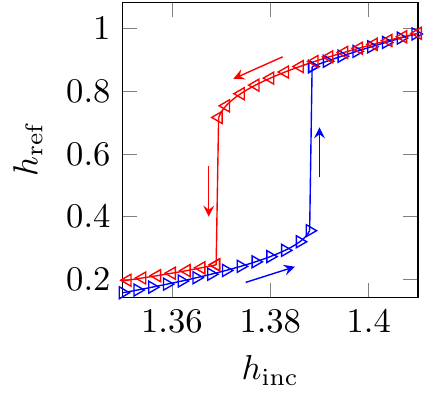}
		\caption{reflected magnetic field amplitude}
	\end{subfigure}%
	\begin{subfigure}[t]{0.5\columnwidth}
	  \centering
		\includegraphics{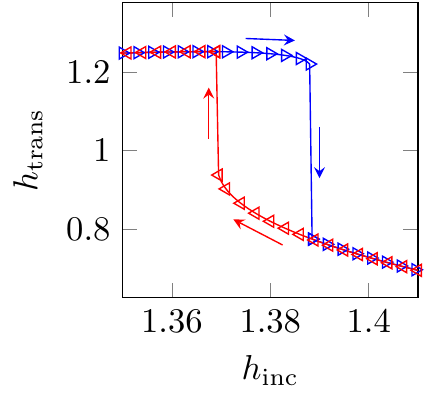}
		\caption{transmitted magnetic field amplitude}
	\end{subfigure}%
	\caption{\label{fig:hysteresis_amp} 
		 Stationary state amplitude of $\varphi$ (a), reflected magnetic field amplitude (b) and transmitted magnetic field amplitude (c) against incoming magnetic field amplitude $h_{\rm inc}$ for $\alpha=0.1$, $\beta=1.5$ and $\omega_{\rm D}=1.14$.
	}
\end{figure}
Figure~\ref{fig:hysteresis_amp}(a) shows simulation results of the first kind, applying a source term according to \eqref{hsweep} to an rf-SQUID with parameters $\alpha=0.1$ and $\beta=1.5$. One can see the amplitude of the stationary state oscillation of $\varphi$, belonging to different incident plane wave amplitudes. The blue triangles pointing to the right indicate dynamic parameter sweep simulation results from small amplitudes upwards towards larger ones. Vice versa, the red triangles pointing to the left indicate dynamic sweep simulation results from large amplitudes downwards towards smaller ones. One can observe, that in a certain range of amplitudes the curves do not coincide. The hysteresis loop that occurs for $\varphi$ can also be observed in the reflected and transmitted field amplitudes, see Fig. \ref{fig:hysteresis_amp}(b)-(c). In a certain bistable region, the amplitudes of the reflected and transmitted waves, respectively, do not only depend on the amplitude of the incident plane wave, but also on the direction this amplitude value has been approached from.

\begin{figure}[t]
	\centering
	\begin{subfigure}[t]{0.5\columnwidth}
	  \centering
		\includegraphics{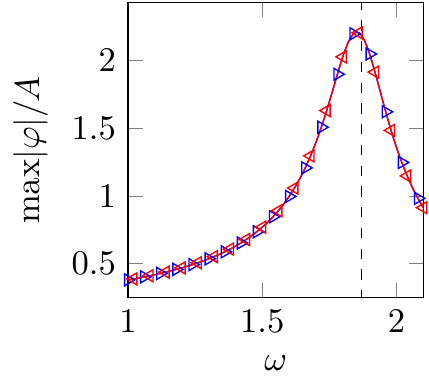}
		\caption{stationary state amplitude for $h_{\rm inc}=  10^{-3}$}
	\end{subfigure}%
	\begin{subfigure}[t]{0.5\columnwidth}
	  \centering
		\includegraphics{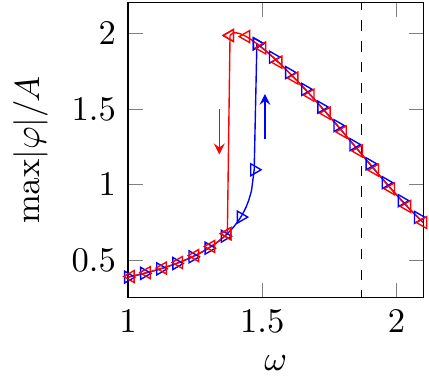}
		\caption{stationary state amplitude for $h_{\rm inc}=1.2$}
	\end{subfigure}
	\begin{subfigure}[t]{0.5\columnwidth}
	  \centering
		\includegraphics{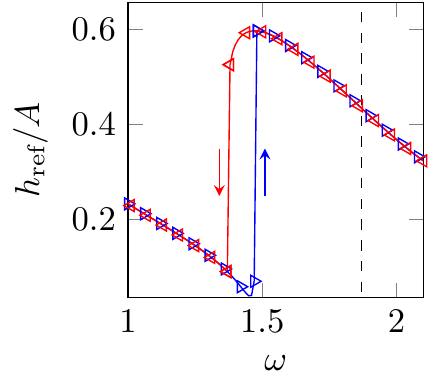}
		\caption{reflected magnetic field amplitude}
	\end{subfigure}%
	\begin{subfigure}[t]{0.5\columnwidth}
	  \centering
		\includegraphics{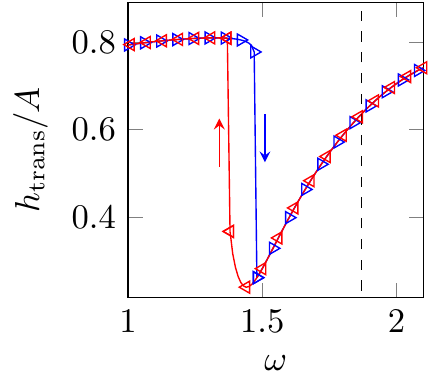}
		\caption{transmitted magnetic field amplitude}
	\end{subfigure}
	\caption{\label{fig:hysteresis_freq}
		For $h_{\rm inc}=  10^{-3}$: stationary state amplitude of $\varphi$ for (a). For  $h_{\rm inc}=1.2$: stationary state amplitude of $\varphi$ (b), reflected field amplitude (c) and transmitted field amplitude (d). All plotted against the frequency $\omega$ for $\alpha=0.1$, $\beta=2.5$.
	}
\end{figure}

We further apply a plane wave source according to \eqref{omegasweep}, i.e., we keep the amplitude of the incident plane wave fixed throughout the entire simulation and sweep its driving frequency $\omega_{\rm D}$ over a frequency range, which contains the resonance frequency $\omega_0=\sqrt{1+\beta}$. The results are shown in Fig. \ref{fig:hysteresis_freq}(b)-(d). The resonance frequency is indicated by the vertical black dashed lines. In Fig. \ref{fig:hysteresis_freq}(a), $h_{\rm inc}=10^{-3}$ is in the linear interaction regime. The other three figures are plots of a simulation done at $h_{\rm inc}=1.2$, when nonlinear effects already play a role. The damping parameter is chosen to be $\alpha=0.1$ and the SQUID parameter is $\beta=2.5$. The sweeps were done first from $\omega_{\rm min}=1.3$ to $\omega_{\rm max}=2.1$ in an increasing way (blue triangles pointing to the right), afterwards vice versa in a decreasing way (red triangles pointing to the left). The response of the system is different in either case, exhibiting the manifestations of the nonlinear terms in the equations of motion. We can observe that the resonance of both the phase difference $\varphi$ and the reflected wave $h_{\rm ref}$ are shifted to smaller frequencies. Compare this observation to the Duffing oscillator, which takes into account the cubic term in the $\sin$-expansion as well \cite{Sharma2012, Wu2014}. Hence, the minimum of the transmitted wave amplitude occurs at smaller frequencies than in the linear case as well. Thus, one can tune the effective resonance frequency of the entire system by increasing the amplitude of the incoming magnetic field amplitude $h_{\rm inc}$ to smaller frequencies.
\newline

\section{\label{sec:realistic values} Discussion of realistic parameters}
In order to retain a physical intuition of the dimensions of the involved quantities, we briefly insert realistic values into the parameters of the normalized system introduced in section \ref{sec:normalization}. From Ref.~\onlinecite{Butz2013} and Ref.~\onlinecite{Butz2015}, we take the values presented in the experimental study of a transmission line based rf-SQUID interaction setup. We plug in the geometric inductance $L=83\,\rm pH$, an intrinsic capacitance of the JJ of $C=0.02\,\rm pF$, a critical current of the JJ of $I_{\rm cr}=1.8\,\rm \upmu A$, and a resistance $R=1600\,\rm \Omega$. This results in the characteristic frequency $\tilde{\omega}\approx 2\pi\cdot 124\,\rm GHz$ and a scaling parameter for the spatial coordinates of $\tilde{\lambda}\approx 0.4\,\rm mm$. Due to the large resistance, the damping parameter is $\alpha\approx 0.04$ and the rf-SQUID in this case is nonhysteretic with $\beta\approx 0.45$. We want to point out here, that in this contribution we used a higher value $\alpha=0.1$ (corresponding to a resistance of $R=640\,\rm \Omega$) to investigate the effect of dissipative losses, which otherwise wouldn't have shown its impact on the dynamics of the system. All other parameters remain unchanged by this replacement of the resistance. We would also like to emphasize that the authors of Ref.~\onlinecite{Butz2013} and Ref.~\onlinecite{Butz2015} used an additional shunted capacitance of $C=2\,\rm pF$ to lower the resonance frequency of the rf-SQUID by roughly one order of magnitude to the range of $\omega\approx 2\pi\cdot 10\,\rm GHz$. In this study, we generically used $\kappa=1$. We obtain this value by assuming $R_{\rm i}=10\,\rm \upmu m$, according to Fig. \ref{fig:geometry}, and a cross sectional area of $A_{\rm c}=1600\,\upmu m^2$. This corresponds to a torus-shaped ring with quite small aspect ratio. The operational frequency is in the range of $f\approx 100\,\rm GHz$. Consequently, the wavelength $\lambda$ is in the order of $\rm mm$. The above choice of geometry is therefore justified and we meet the condition $d\ll\lambda$, since the wavelength is two orders of magnitude larger than the radius of the rings. The scaling parameter for the magnetic field evaluates to $\vec{h}=1.2\,\frac{\rm m}{\rm A}\cdot\vec{H}$. This means, that a value of $h=1$ corresponds to a magnetic induction of $B=1.05\,\rm \upmu T$.

\section{\label{sec:Conclusion} Conclusion}

We have derived an interaction model of an rf-SQUID loaded infinitesimally thin film with electromagnetic waves. In a strictly mathematical treatment, we showed that our problem is wellposed. Therefore, a unique solution of the system of coupled differential equations exists. 

We have treated the model in the linear small-amplitude interaction regime analytically. In this limit, we derived analytical expressions for the reflection and transmission coefficients of the film as well as for the absorption function.

To investigate nonlinear effects, we proposed a numerical scheme based on the finite element method and the Crank-Nicolson scheme. We further showed rigorous error estimates and presented a numerical scheme based on transparent boundary conditions with inflow, where the computational effort is independent of the spatial resolution. With these methods, we simulated the dynamics in the system numerically and found bistable and hysteretic behavior in the nonlinear interaction regime.

In future work, interaction between the rf-SQUIDs inside the film has to be taken into account. It has been proposed already to assume an interaction via mutual inductance between the rings \cite{Hizanidis2016, Lazarides2015, Lazarides2013}. Moreover, one has to investigate, if the coupling of the electric component of the wave to the rf-SQUID is relevant in the description of the interaction.  

\section*{Acknowledgements}
	The authors gratefully acknowledge financial support by the Deutsche Forschungsgemeinschaft (DFG) through CRC 1173. We also wish to thank Jan Brehm and Alexey Ustinov for many helpful comments.

\bibliography{references}

\end{document}

%% file: texdef.tex
\usepackage{graphicx}
\usepackage{dcolumn}
\usepackage{bm}
\usepackage{mathtools}          
\usepackage[english]{babel}
\usepackage{ragged2e}
\usepackage{ifthen}             

\usepackage{pgfplots}           
\usepackage{subcaption}

\newcommand{\R}{\mathbb{R}} 
\newcommand{\N}{\mathbb{N}} 

\newcommand{\X}{X}
\newcommand{\V}{V}
\newcommand{\mt}{m}
\newcommand{\bt}{b}
\newcommand{\at}{a}
\newcommand{\ft}{f}

\newcommand{\Xo}{X_1}
\newcommand{\Vo}{V_1}
\newcommand{\mo}{m}
\newcommand{\bo}{b}
\renewcommand{\ao}{a}
\newcommand{\fo}{f}
\renewcommand{\u}[1][]{
    \ifthenelse{\equal{#1}{}}{u}{}
    \ifthenelse{\equal{#1}{1}}{\vec{u}_h}{}
    \ifthenelse{\equal{#1}{2}}{u_\varphi}{}
}
\renewcommand{\v}[1][]{
    \ifthenelse{\equal{#1}{}}{v}{}
    \ifthenelse{\equal{#1}{1}}{\vec{v}_h}{}
    \ifthenelse{\equal{#1}{2}}{v_\varphi}{}
}
\newcommand{\w}[1][]{
    \ifthenelse{\equal{#1}{}}{w}{}
    \ifthenelse{\equal{#1}{1}}{\vec{w}_h}{}
    \ifthenelse{\equal{#1}{2}}{w_\varphi}{}
}

\renewcommand{\H}{{\vec{h}}}
\newcommand{\uo}{\u_0}
\newcommand{\Ho}{\H_0}

\newcommand{\upo}{u_{\texttt{t},0}}
\newcommand{\Hpo}{\vec{h}_{\texttt{t},0}}

\newcommand{\n}{\vec{n}} 
\newcommand{\Hin}{\vec{h}_{\textnormal{in}}} 

\newcommand{\dx}{\,dx}

\newcommand{\Ol}{\Omega_\ell}

\newcommand{\Pk}{\mathcal{P}_k}
\newcommand{\Voh}{\widehat{V}}
\newcommand{\uh}{\widehat{\u}}
\newcommand{\uhp}{\widehat{\u}_{\texttt{t}}}
\newcommand{\uhn}{\widehat{\u}^n}

\newcommand{\uho}{\uh_{0}}
\newcommand{\vh}{\widehat{v}}
\newcommand{\uhpn}{\uhp^n}
\newcommand{\uhpo}{\widehat{\u}_{\texttt{t},0}}
\newcommand{\Olh}{\widehat{\Omega}_{\ell}}
\newcommand{\moh}{\widehat{m}}
\newcommand{\aoh}{\widehat{a}}

\newcommand{\Id}{I}
\newcommand{\Ah}{\widehat{A}}
\newcommand{\Bh}{\widehat{B}}

\DeclarePairedDelimiter{\norm}{\lVert}{\rVert}

\newcommand{\tn}{t_n}

\newlength\LineWidth
\newlength\MarkSize